\documentclass[journal]{IEEEtran}

\usepackage{xcolor,soul,framed} 

\colorlet{shadecolor}{yellow}
\usepackage[pdftex]{graphicx}
\graphicspath{{../pdf/}{../jpeg/}}
\DeclareGraphicsExtensions{.pdf,.jpeg,.png}

\usepackage[cmex10]{amsmath}
\usepackage{array}
\usepackage{mdwmath}
\usepackage{mdwtab}
\usepackage{eqparbox}
\usepackage{url}
\usepackage{graphicx}

\usepackage{amssymb}
\usepackage{booktabs}
\usepackage{multirow}
\usepackage[T1]{fontenc}
\usepackage{comment}

\begin{document}
\bstctlcite{IEEEexample:BSTcontrol}

\title{ IoT-based Smart Real-Time Energy-Aware Server Room Monitoring System}

\author{
Muhammad Dikko Gambo,
Sami Ul Haq,
Syed M. Minhaj Naqvi,
Khaled M. Rabie,~\IEEEmembership{Senior Member,~IEEE},
and Abdallah Moubayed,~\IEEEmembership{Senior Member,~IEEE}%
\thanks{
Muhammad Dikko Gambo, Sami Ul Haq, and Syed M. Minhaj Naqvi are with the Department of Computer Engineering, King Fahd University of Petroleum \& Minerals (KFUPM), Dhahran 31261, Saudi Arabia (e-mail: g202304070@kfupm.edu.sa; g202315350@kfupm.edu.sa; g202426160@kfupm.edu.sa).
}%
\thanks{
Khaled M. Rabie is with the Department of Computer Engineering and the Research Center for Communication Systems and Sensing, King Fahd University of Petroleum \& Minerals (KFUPM), Dhahran 31261, Saudi Arabia (e-mail: k.rabie@kfupm.edu.sa).
}%
\thanks{
Abdallah Moubayed is with the Department of Computer Engineering and the Interdisciplinary Research Center for Intelligent Secure Systems, King Fahd University of Petroleum \& Minerals (KFUPM), Dhahran 31261, Saudi Arabia (e-mail: abdallah.moubayed@kfupm.edu.sa).
}%
\thanks{Manuscript received XX XX, 2026.}
}



\maketitle

\begin{abstract}
The integration of the Internet of Things (IoT) into monitoring systems has become increasingly significant across various fields, including energy management, environmental monitoring, and industrial applications. This paper proposes a smart, real-time, energy-aware monitoring system architecture for IoT using a brokerless Data Distribution Service (DDS) middleware. Environmental and electrical metrics are transmitted with high reliability, achieving an average latency of 1.76 milliseconds and a throughput of 57.1 Mbps. These metrics are processed by a multi-head Long Short-Term Memory network. Evaluated on the empirical TDC2.0 data center dataset, the architecture simultaneously performs three tasks: predicting thermal stress, detecting cooling inefficiency, and forecasting multi-step cooling power demand. The multi task network achieved an F1-score of 0.92 for proactive thermal stress prediction and demonstrated the capability to identify systemic cooling anomalies. Furthermore, the model generated consistent power forecasts across 5-minute, 10-minute, and 30 minute horizons, maintaining an R2 > 0.83. A post-hoc SHAP analysis indicated that the predictive logic aligns with established thermodynamic relationships. Overall, the proposed architecture offers a transparent and low-latency framework for critical facility management.
\end{abstract}



%
\IEEEpeerreviewmaketitle


\section{Introduction}
The Internet of Things (IoT) has significantly advanced industrial and residential energy management by enabling real-time data collection and analysis. IoT-based monitoring systems integrate distributed sensors with communication technologies to monitor equipment operating conditions continuously. This capability supports anomaly detection, early identification of operational deviations, and timely maintenance interventions \cite{9111678}. Some of the most energy-intensive components on the power grid include elements of modern digital infrastructure like data centers and server rooms, with recent studies estimating that the data centres account for almost 1-2\% of global electric consumption, and this figure is expected to rise with increasing dependence on cloud computing, artificial intelligence, and large scale computational workloads \cite{iea_energy_ai_report} \cite{iea_ai_energy_2024}. A big part of this consumption constitutes not the IT hardware itself, but the equipment related to cooling systems and other climate control appliances, so as to maintain safe and secure operating conditions. Cooling alone accounts for up to 40\% of the energy used, strengthening the role of cooling efficiency as a critical factor when it comes to sustainability and cost management \cite{doe_cooling_2023} \cite{reuters_datacenter_cooling_2025}. Optimizing energy use in server rooms aligns with the United Nations Sustainable Development Goals (SDGs), particularly SDG 7: Affordable and Clean Energy and SDG 9: Industry, Innovation, and Infrastructure, by promoting energy-efficient operations and innovative monitoring systems that reduce environmental impact. The main job of server rooms is to host essential IT infrastructure including servers, network appliances, digital storage equipment, and other auxiliary systems. Ensuring optimal levels of temperature and humidity plays a vital role in getting reliability, availability, and long lifetime for such hardware \cite{ashrae_thermal_guidelines} \cite{lbl_server_data_2021}. Deployment of environment monitoring system is a common solution for tracking these important parameters and get alerts as set thresholds are exceeded.

A number of studies have articulated usage of IoT and wireless sensor networks for monitoring of environment parameters. As an example, Anwar et al. developed a web-based system integrating microcontrollers and sensors for temperature and humidity tracking in server rooms \cite{anwar2017monitor}. On the other hand, Yumang et al. put forth the idea of using a ZigBee based monitoring framework for server room environmental monitoring \cite{yumang2016zigbee}, while Rani et al. preferred using Message Queuing Telemetry Transport (MQTT) based solutions for delivering real-time environment data to administrators \cite{rani2024mqtt}, where MQTT is a lightweight publish-subscribe messaging protocol. Though useful, the scope of these studies is stunted to only provide insights into environmental conditions, leaving out any discussions on energy efficiency as it pertains to cooling systems which have the propensity to consume extensive electrical power even as environmental parameters are within acceptable ranges. This results in wastage of unnecessary operational cost and energy. As the emphasis placed on sustainability and efficiency is growing, the need for systems that not only capture the environmental data, but also provide analysis on cooling energy usage, is highly pressing. Advancements in smart metering and machine learning (ML) technologies offer promising opportunities to address this issue. Smart meters enable remote monitoring of electrical power consumption. Moreover, ML solutions have gotten efficient at understanding obscure patterns in environment and energy data and make out inefficiencies, narrow down anomalies, and also predicting optimal operating conditions \cite{google_datacenter_efficiency} \cite{IEA2025EnergyAI}. Leveraging these technologies, monitoring systems can evolve from passive sensing tools into intelligent decision-support systems that optimize energy use proactively. 

These aforementioned challenges motivate us to propose a smart and real time energy aware server room monitoring system, integrating environment sensing, smart energy metering, and ML techniques to deliver an intelligent monitoring framework, which is capable of detecting inefficient cooling and also support energy oriented decision making in these big IT facilities. We aim to optimize cooling energy consumption while maintaining safe environmental conditions for server rooms. 

The key contributions of this research are summarized as follows:
\begin{itemize}
    \item We propose a decentralized, real-time server room monitoring architecture that decouples edge sensing from central analytics using a brokerless Data Distribution Service (DDS), addressing the scalability and latency limitations of traditional centralized approaches.
    \item We design a multi-head Long Short-Term Memory (LSTM) network capable of simultaneously performing three interrelated predictive tasks based on combined environmental and electrical telemetry: proactively detecting thermal stress, identifying cooling inefficiency, and forecasting multi-step cooling power demand.
    \item We propose effective, efficient DDS  Quality of Service (QoS) policies to ensure real-time performance in server-room monitoring. 
    
    \item We evaluate the system-level performance, demonstrating deterministic low latency and high-throughput data delivery. 
    \item We validate the ML model on the real-world TDC2.0 operational dataset, demonstrating satisfactory multi-task performance. Furthermore, we provide a post-hoc SHAP interpretability analysis to confirm that the network's predictive logic aligns with sound thermodynamic principles.
\end{itemize}

\section{Literature Review}
Early work focused on the importance of the thermal visibility itself. Wang et al. proposed a computational fluid dynamics based sensor placement method for hot server detection in the data centers, exhibiting that effective monitoring is a factor of not only the sensing hardware but also of strategic sensor deployment \cite{wang2013sensor}. In the same vein, Mousavi et al. studied server room cooling through simulation and demonstrated that thermal models can be utilized for exploration of energy-intelligent control strategies that balance processor thermal comfort with reduced cooling energy use \cite{mousavi2015energy}. These studies established the importance of thermal monitoring, but they absent integration of direct power metering or of real-time intelligent analytics into a unified monitoring framework.

With the rise of the low-cost embedded technologies and IoT technologies, several work shifted toward practical server room monitoring and its control. Purwanto et al. developed a microcontroller based temperature and humidity control system that used fuzzy logic to adjust air conditioner settings automatically \cite{purwanto2018fuzzy}. Onibonoje et al. build on this idea by proposing an IoT based server room conditioning system capable of monitoring heat, humidity, smoke, flame, water leakage, and power outages, while also enacting corrective measures and sending cloud and mobile notifications in real time \cite{onibonoje2019iot}. These works show that the IoT-based server room supervision has potential to improve environmental safety and automation. However, they mainly focus on maintaining acceptable conditions and alarm generation, rather than determining whether cooling is being performed efficiently from an energy use perspective.

More recent work address energy-aware cooling optimization in data centers through ML and advanced control. Li et al. put forth cooling optimization as a deep reinforcement learning problem and showed that learning-based control has the potential to reduce cooling energy cost while preserving thermal safety conditions \cite{li2020drl}. Zhang et al. provided a comprehensive survey of data-center cooling technologies, power-consumption modeling, and optimization strategies, ultimately highlighting how cooling remains one of the most critical targets for energy savings in digital infrastructure \cite{zhang2021survey}. Building on the same direction, Chen et al. used machine-learning algorithms to predict optimal operating parameters for a centralized data-center heat-dissipation system and reported potential improvements in Power Usage Effectiveness (PUE) \cite{chen2023mlcooling}. Li et al. further combined physical laws with artificial neural networks to improve prediction accuracy for cooling-system energy consumption, illustrating the value of hybrid physics-informed learning models \cite{li2024physicsann}. Even more recently, Kula et al. demonstrated that machine-learning forecasting of temperature in a data-center cooling system has potential to support more intelligent cooling adjustment in real operating environments \cite{kula2025mltemp}.

Another relevant direction concerns real-time middleware and intelligent energy telemetry. The study most closely related to the communication and analytics aspects of the present work is that of Shuaibu et al., who proposed a smart metering system that combines high-resolution electrical sensing, Fast DDS middleware, and deep-learning-based anomaly detection \cite{shuaibu2026smart}. This work is not specifically targeted at server-room cooling, but it demonstrates the value of broker-less publish–subscribe communication for low-latency telemetry and AI-assisted energy monitoring. This makes it particularly relevant to the present study, which also requires dependable real-time data exchange alongside intelligent analysis.  
Table \ref{tab:comparison} summarizes the aspects tackled in this work relative to other previous works from the literature. 

\begin{table*}[htbp]
\centering
\caption{Comparison of Related Works ($\checkmark$ = present, $\times$ = not present, $\triangle$ = partial, --- = not applicable)}
\label{tab:comparison}

\begin{tabular}{|c|c|c|c|c|c|}
\hline
\textbf{Study} & \textbf{Env. Monitoring} & \textbf{Power Metering} & \textbf{ML/AI} & \textbf{Real-Time Platform} & \textbf{DDS/QoS} \\
\hline
\cite{wang2013sensor}  & $\checkmark$ & $\times$ & $\times$ & $\times$ & $\times$ \\
\cite{mousavi2015energy}  & $\triangle$  & $\times$ & $\times$ & $\times$ & $\times$ \\
\cite{purwanto2018fuzzy}  & $\checkmark$ & $\times$ & $\triangle$ & $\checkmark$ & $\times$ \\
\cite{onibonoje2019iot}  & $\checkmark$ & $\times$ & $\times$ & $\checkmark$ & $\times$ \\
\cite{li2020drl}  & $\triangle$  & $\times$ & $\checkmark$ & $\times$ & $\times$ \\
\cite{zhang2021survey}  & ---          & ---      & ---      & $\times$ & $\times$ \\
\cite{chen2023mlcooling}  & $\triangle$  & $\times$ & $\checkmark$ & $\triangle$ & $\times$ \\
\cite{li2024physicsann}  & $\triangle$  & $\times$ & $\checkmark$ & $\times$ & $\times$ \\
\cite{kula2025mltemp}  & $\checkmark$ & $\triangle$ & $\checkmark$ & $\triangle$ & $\times$ \\
\cite{shuaibu2026smart} & $\times$ & $\checkmark$ & $\checkmark$ & $\checkmark$ & $\checkmark$ \\
\hline
\textbf{This Work} & $\checkmark$ & $\checkmark$ & $\checkmark$ & $\checkmark$ & $\checkmark$ \\
\hline
\end{tabular}
\end{table*}

Overall, the literature reveals a clear gap. Existing server-room studies mainly emphasize environmental monitoring and rule-based or fuzzy control \cite{purwanto2018fuzzy}\cite{onibonoje2019iot}, whereas data-center energy studies concentrate on cooling optimization, prediction, or learning-based control at broader facility scale. Meanwhile, DDS-based intelligent metering has demonstrated the promise of real-time distributed energy analytics, but it has not been directly adapted to server-room monitoring and cooling-efficiency assessment \cite{shuaibu2026smart}. Therefore, there remains a need for a unified system that combines environmental sensing, direct cooling-energy metering, machine-learning-based detection of unsafe and inefficient operating conditions, and real-time DDS QoS mechanisms in a server-room setting. The proposed work is intended to address this gap.  

\section{METHODOLOGY}
\begin{figure}[htbp]
    \centering
    \includegraphics[width=\columnwidth]{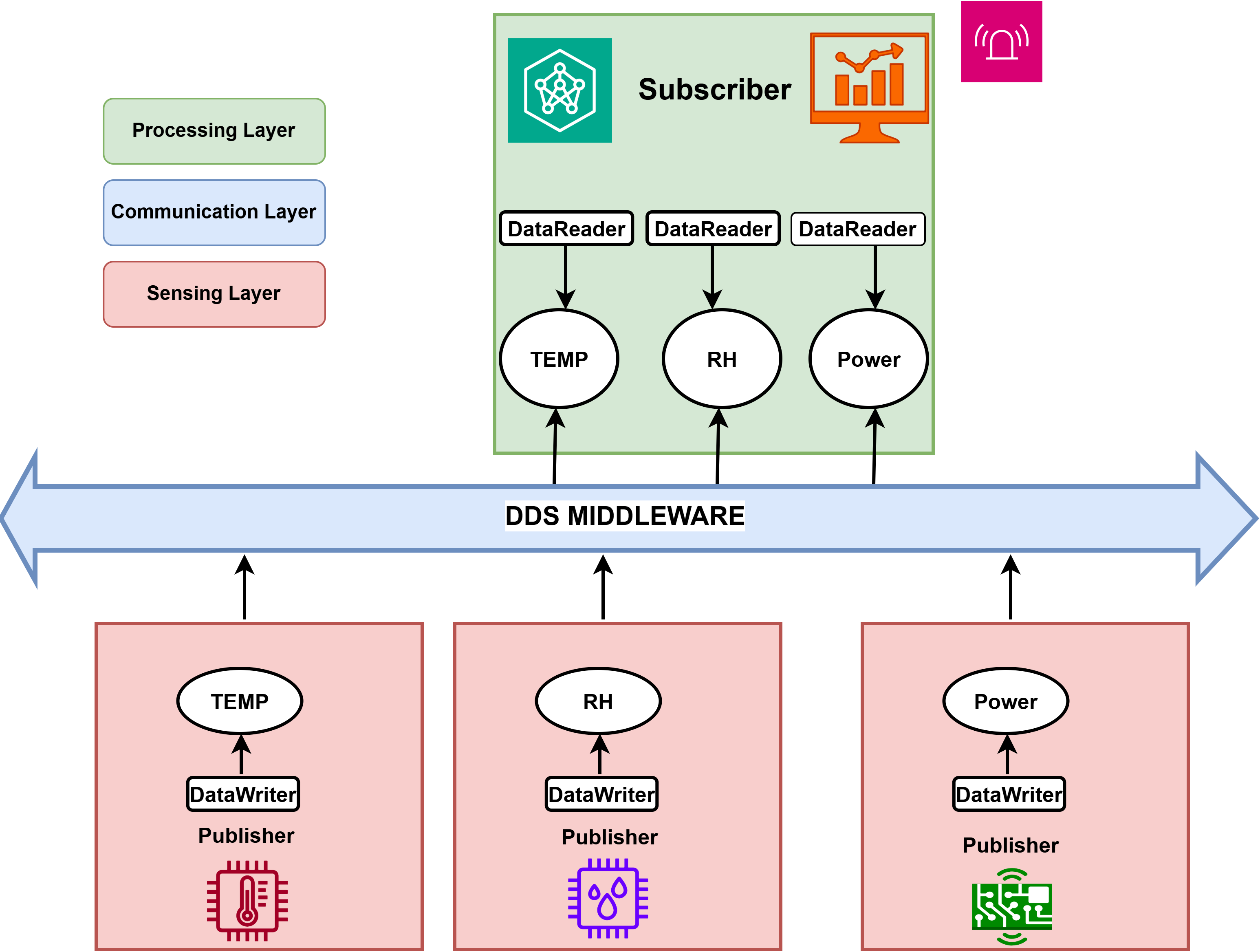}
    \caption{ Proposed system architecture with three layers: sensing, communication, and processing.}
    \label{fig:proposed-system}
\end{figure}

\subsection{The Proposed System}



The proposed system's architecture is shown in Figure \ref{fig:proposed-system}. The system is logically split into three layers: the Sensing Layer, the Communication Layer, and the Processing Layer. At the edge, environmental and electrical metrics are collected at the same time and sent through a brokerless publish-subscribe middleware before it is finally used by centralized computational units for feature engineering and multi-task ML inference. Below are the specific roles of each layer.

\subsubsection{Sensing Layer}


This layer comprises three distributed publisher nodes: a temperature sensor, a humidity sensor, and a smart meter associated with the cooling unit. Each node periodically publishes timestamped measurements together with sensor identity and location. The environmental sensors provide visibility into server-room operating conditions, while the smart meter provides electrical measurements such as voltage, current, power, frequency, and power factor. This combination enables the system to monitor both environmental safety and cooling-energy usage.

\subsubsection{Communication Layer}

This layer is implemented using  DDS which provides a data-centric publish/subscribe model with automatic discovery, direct peer-to-peer communication, and configurable QoS. These properties make it suitable for distributed real-time monitoring systems. In the proposed design, all sensing nodes and the server communicate through typed DDS Topics within a common domain. The DDS provides support for low-latency data flow, reliable messaging, deadline monitoring, and loosely coupled distributed applications.
For telemetry delivery, the sensing streams use QoS settings appropriate for periodic monitoring data.  The selected communication behavior uses reliable reliability, keep-last history, transient-local durability, and deadline-aware communication, ensuring dependable delivery of the latest valid sample while allowing timely detection of missing updates. 

\subsubsection{Processing Layer}

This layer is hosted on a central server and performs data aggregation, monitoring, and intelligent analysis. It subscribes to the {temperature}, {humidity}, and {smart-meter} topics and maintains the latest view of the monitored environment.


Rather than relying on traditional static alarms that only trigger after a safety threshold has already been breached, this layer is dedicated entirely to proactive, ML based predictive analytics. Specifically, we trained a multi-head LSTM model capable of analyzing the latent temporal patterns. The network simultaneously generates three predictive outputs:

\begin{enumerate}
    \item Prediction of future unsafe environmental conditions,
    \item Prediction of cooling inefficiency, and
    \item Forecasting of future cooling power demand.
\end{enumerate}


 By leveraging the correlations between environmental heat loads and mechanical electrical consumption, the processing layer provides facility operators with advanced early warnings. This enables the identification of impending thermal stress and systemic cooling inefficiencies well before they manifest as critical hardware risks or severe energy waste.

\subsection{Proposed Multi-Task ML-based Architecture}

\begin{figure*}[htbp]
    \centering
    \includegraphics[width=\textwidth]{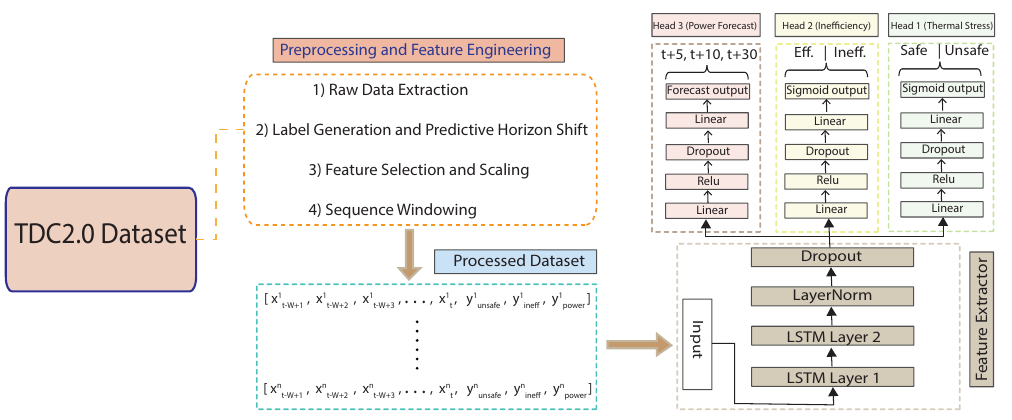}
    \caption{Overall architecture of the proposed multi-task machine-learning model.}
    \label{fig:architecture}
\end{figure*}
The overall architecture of the proposed multi-task machine-learning based model is illustrated in Fig. \ref{fig:architecture}. The methodology is structured as a continuous end-to-end pipeline, beginning with the extraction and preprocessing of raw sensor data from the data center environment. The data is subsequently windowed into temporal sequences and fed into a shared multi-head LSTM network, which is jointly optimized to perform simultaneous prediction of unsafe environmental conditions, cooling inefficiency, and multi-step cooling-power forecasting.

\subsubsection{Data Preprocessing and Feature Engineering}


The pre-processing process can be broken down into four processing steps: raw data extraction (from the dataset files of TDC2.0), predictive label generation, feature selection and scaling, and finally sequence windowing. The primary source for our data is the Air Cooled Tropical Data Center Dataset \cite{N9/BLBQ2T_2023} which includes all of the operational metrics (i.e., power consumption, server temperatures, etc.) from a live data center located in Singapore, utilizing a direct expansion cooling system with 521 active servers running actual production workloads. The dataset includes detailed physical measurements taken at 30-second intervals over a six-month period, from September 2022 to February 2023, that capture both energy use and ambient variables.

\textbf{I) Raw Data Extraction:}
The raw dataset is made up of daily records that are spread out across several sensor gateways. For our monitoring goals, we focus on SensorGW 2, which keeps track of the environmental temperature and humidity (TH) readings. This gateway specifically records the weather outside, which includes the temperature and humidity of the `Outdoor Air' (OATH). It also records the temperature and humidity of the `returned hot air' from the IT racks to the cooling units (RATH), as well as the `supplied cold air' that the cooling units blow into the server room (SATH). We also use SensorGW 5, which keeps track of how much electricity each part of the cooling unit uses. More precisely, facility's cooling system consists of two precision cooling units, that is PCU 4 and PCU 5. Each PCU consist of these power consumption parts: the CRAC Main (PCM), compressor 1 and compressor 2 (PCP1 and PCP2), the evaporator fan (PEF), and the condenser fan (PCF). 


The data extracting and aligning pipeline processes the daily records by combining the environmental and power readings based on their timestamps. Apart from the features in the datasets, two derived features are calculated to explicitly quantify the different heat sources in the data center.

First, the weather load delta ($\Delta T_{\text{weather}}$), which measures the difference between the outside weather and the cold supply air injected into the room:
\begin{equation}
\Delta T_{\text{weather}} = T_{\text{OATH}} - T_{\text{SATH.}}
\end{equation}
This derived metric shows how much cooling is needed to beat the heat outside. It helps the model separate external weather stress from activity on the server.

Second, the internal heat load delta ($\Delta T_{\text{internal}}$), which measures the difference between the hot return air from the servers and the cold supply air:
\begin{equation}
\Delta T_{\text{internal}} = T_{\text{RATH}} - T_{\text{SATH.}}
\end{equation}

This metric indicates how much heat the computer equipment adds to the air. Calculating this delta is a necessary preprocessing step because it is the basis for defining the cooling inefficiency labels. However, it is ultimately left out of the final model inputs to protect against data leakage.

For each precision cooling unit, the deltas for temperature and humidity are calculated separately. Finally, the total cooling power for the whole facility is found by adding up the readings from each component.

\textbf{II) Label Generation and Predictive Horizon Shift:}
Three different predictive targets are created for the multitask learning framework to transform the model from a mere description of the present state into an active forecast of the future state. These consist of two binary classification labels and one continuous regression target. The ground truth values of all three targets are moved forward in time with respect to the input sensor readings so that the model can forecast.

\textbf{Label 1 (Thermal Stress):} This label is meant to serve as an early warning mechanism, whereby the operators of the facilities can take action before the server room becomes dangerously hot. We define thermally stressed conditions as the top 15th percentile of return air temperature across the entire dataset for both PCUs at any given timestep:
\begin{equation}
\text{RATH}_{\max}(t) = \max(\text{RATH1}_{\text{temp}}(t), \text{RATH2}_{\text{temp}}(t))
\end{equation}
\begin{equation}
y_{\text{unsafe}}(t) = 
\begin{cases} 
1, & \text{if } \text{RATH}_{\max}(t) \geq Q_{85}(\text{RATH}_{\max}) \\ 
0, & \text{otherwise,} 
\end{cases}
\end{equation}
where $Q_{85}$ denotes the 85th percentile. This means that the top 15\% of all the `return air temperature' readings across the whole dataset are labeled ``Stressed''. To make this task predictive, this binary label is shifted forward by 20 timesteps. This teaches the model to identify the hidden sensor patterns that precede a thermal event by exactly 10 minutes.

\textbf{Label 2 (Cooling Inefficiency):} The purpose of this label is to proactively flag wasteful operating windows where the cooling units consume excessive electrical energy without removing a proportionate amount of heat. To quantify this, a Coefficient of Performance (COP) proxy is constructed as the ratio of heat removed to electrical energy consumed:
\begin{equation}
\text{COP}_{\text{proxy}} = \frac{\Delta T_{\text{internal}}}{\text{Total\_Power\_W}}
.\end{equation}
Timesteps falling below the 33rd percentile of this proxy are labelled as inefficient ($y_{\text{ineff}} = 1$). The 33rd percentile is computed across the entire dataset. Similar to the thermal stress label, this target is shifted forward by 10 minutes so the model learns to anticipate upcoming drops in system efficiency.

\textbf{Label 3 (Power Forecast):} Unlike the first two classification tasks, the purpose of this third target is to provide a continuous regression forecast of the total facility cooling power. This provides grid operators and facility managers with the expected energy load, aiding in power allocation. The raw facility cooling power values are shifted forward by 10, 20, and 60 timesteps, creating three simultaneous regression targets that represent the 5 minute, 10 minute, and 30 minute future horizons, respectively.

\textbf{III) {Feature Selection and Scaling}:}
To maintain the predictive integrity of the model, direct indicators of the target variables must be omitted from the input space to avert data leakage. The thermal stress label comes directly from the RATH, so adding RATH or the internal heat load deltas as input features would let the network easily guess the future thermal states based on the current conditions, as temperature change slowly. As a result, the model is not able to learn how the system works.  By not including these downstream variables, the model is forced to predict future thermal states only from upstream operational signals.

The final feature space has $N_f = 25$ variables, which are logically grouped into four physical categories. The environmental category has 8 features, including the temperature and humidity readings for the outside air and the supply air in the two precision cooling units. Ten parameters belong to the electric domain since they provide information about the power consumption of five specific equipment (CRAC main, compressor 1 and 2, evaporator fan and condenser fan) for the two cooling units. The derived category has 4 features which represent the temperature/humidity weather load delta calculated for each of the two units. Finally, the system totals add three new features: they show the total power consumption for each unit as well as the total for the whole facility.

Because neural networks are highly sensitive to the varying magnitudes of raw physical measurements, all 25 features are independently scaled to a continuous range of $[0, 1]$ using Min Max normalization. Crucially, separate scaling transformations are fitted for the input features and the regression targets. Maintaining a dedicated target scaler ensures that the continuous power forecasts generated by the third predictive head can be accurately inverse transformed back into real watt values during evaluation. This step is mandatory for computing physically interpretable error metrics. The precise dimensions and feature composition of the final processed dataset are detailed subsequently in Section \ref{sec:experimental_setup}.

\textbf{IV) {Sequence Windowing}:}
To capture the thermal inertia and delayed responses characteristic of HVAC systems, the continuous data stream is transformed into sliding windows as the final stage of preprocessing. At any given timestep $t$, the input matrix $X_t \in \mathbb{R}^{W \times N_f}$ is defined as:
\begin{equation}
X_t = [x_{t-W+1}, x_{t-W+2}, \dots, x_t]^T,
\end{equation}
where $W$ represents the window size and $x$ denotes the 25-dimensional scaled feature vector. By applying a stride of 1, the data is organized into $W$-sample sequences, which provide $W$/2 minutes of historical context, as each sample represents 30 seconds.

To formulate the supervised multi-task learning problem, each historical sequence window is explicitly mapped to its corresponding future targets to construct the final dataset matrix. Let $n$ be the total number of generated windows. For any given $i$-th training instance (where $i \in \{1, 2, \dots, n\}$), the complete row vector is formulated as:
\begin{equation}
[x^i_{t-W+1}, x^i_{t-W+2}, \dots, x^i_t, y^i_{\text{unsafe}}, y^i_{\text{ineff}}, y^i_{\text{power}}],
\end{equation}
where the superscript $i$ explicitly denotes the specific window index within the overall dataset. The target variables within this row already incorporate the necessary forward time shifts established during the label generation phase. This structured matrix configuration prepares the data for the shared recurrent backbone.

\subsubsection{Proposed Multi-Head LSTM Architecture}


We propose a multi-task deep learning model, which is based on a LSTM network, to concurrently track thermal stress, identify cooling inefficiency, and predict power consumption. As depicted above in Fig. \ref{fig:architecture}, the model uses a common recurrent backbone to learn common temporal dynamics in the preprocessed sensor data, which is then divided into three task-specific fully connected heads. The layer-by-layer parameters configurations are described in Table \ref{tab:architecture}.

\textbf{I) Shared Recurrent Backbone:}
The core of the architecture is the LSTM designed to process sequential operational data. For a given timestep $k$ within an input window, the operations of an LSTM cell are defined mathematically as follows:
\begin{align}
    f_k &= \sigma(W_f \cdot [h_{k-1}, x_k] + b_f) \\
    i_k &= \sigma(W_i \cdot [h_{k-1}, x_k] + b_i) \\
    \tilde{C}_k &= \tanh(W_C \cdot [h_{k-1}, x_k] + b_C) \\
    C_k &= f_k \odot C_{k-1} + i_k \odot \tilde{C}_k \\
    o_k &= \sigma(W_o \cdot [h_{k-1}, x_k] + b_o) \\
    h_k &= o_k \odot \tanh(C_k),
\end{align}
where $x_k \in \mathbb{R}^{25}$ is the input feature vector, $h_{k-1}$ is the previous hidden state, $C_k$ is the cell state, and $\sigma$ denotes the sigmoid activation function. The matrices $W$ and vectors $b$ represent the learnable weights and biases for the forget gate $f_k$, input gate $i_k$, cell state modulation $\tilde{C}_k$, and output gate $o_k$. The operator $\odot$ denotes element-wise multiplication.

Our backbone utilizes two stacked LSTM layers with a hidden state dimensionality of $d_h = 128$. To summarize the temporal context of the entire window historical sequence, we extract the hidden state strictly from the final timestep. Layer normalization is applied to this state to stabilize the training process by smoothing the loss landscape, followed by a dropout layer with a probability of $p=0.4$ to mitigate overfitting. This produces the final shared representation vector $z \in \mathbb{R}^{128}$:
\begin{equation}
z = \text{Dropout}(\text{LayerNorm}(h_T)).
\end{equation}

\textbf{II) Task Specific Projection Heads:}
The shared representation vector $z \in \mathbb{R}^{128}$ is subsequently routed to three independent feedforward networks. Utilizing a shared backbone forces the model to learn generalized, robust physical features that are beneficial across all three interrelated thermal and electrical tasks, while the independent heads map these generalized features to specific target distributions.

\textbf{Head 1 (Thermal Stress):} This binary classification head is designed to predict the probability of future thermal stress. It consists of a fully connected dense layer that projects the 128 dimensional shared state down to a 32 dimensional intermediate representation. Nonlinear transformations are applied via a Rectified Linear Unit (ReLU) activation, followed by a dropout layer ($p=0.4$) for regularization. A final linear projection collapses the vector into a single unnormalized logit $l_{\text{stress}}$:
\begin{equation}
l_{\text{stress}} = W_{s2}(\text{Dropout}(\text{ReLU}(W_{s1}z + b_{s1}))) + b_{s2}.
\end{equation}
To clarify the mathematical notation, the subscript $s$ explicitly denotes parameters belonging to the thermal stress head. The terms $W_{s1} \in \mathbb{R}^{32 \times 128}$ and $b_{s1} \in \mathbb{R}^{32}$ represent the learnable weight matrix and bias vector of the first dense layer. The terms $W_{s2} \in \mathbb{R}^{1 \times 32}$ and $b_{s2} \in \mathbb{R}$ represent the weight matrix and bias scalar of the final output layer, respectively. The predicted probability of thermal stress, $\hat{p}_{\text{stress}}$, is obtained by applying the sigmoid activation function: $\hat{p}_{\text{stress}} = \sigma(l_{\text{stress}})$.

\textbf{Head 2 (Inefficiency Detection):} Structurally identical to the thermal stress head, this branch maps the shared representation $z$ to a single logit $l_{\text{ineff}}$ to predict future cooling inefficiency based on the COP proxy. It utilizes its own independent weight matrices and bias vectors ($W_{i1}$, $b_{i1}$, $W_{i2}$, $b_{i2}$) to isolate the specific electrical signatures of inefficient equipment operation from the shared temporal context. The predicted inefficiency probability is $\hat{p}_{\text{ineff}} = \sigma(l_{\text{ineff}})$.

\textbf{Head 3 (Power Forecast):} The multi step regression head shares the same 128 to 32 dimensional intermediate projection and activation structure as the classification branches. However, its final linear layer does not employ a sigmoid activation. Instead, it projects the 32 dimensional intermediate vector into three continuous outputs:
\begin{equation}
\hat{y}_{\text{power}} = W_{f2}(\text{Dropout}(\text{ReLU}(W_{f1}z + b_{f1}))) + b_{f2}
\end{equation}
Here, the subscript $f$ denotes the forecasting head. The final weight matrix $W_{f2} \in \mathbb{R}^{3 \times 32}$ projects the data into a three element vector $\hat{y}_{\text{power}} \in \mathbb{R}^3$. These three outputs correspond directly to the predicted normalized facility cooling power at the 5 minute, 10 minute, and 30 minute future horizons.

\begin{table}[htbp]
\centering
\caption{Proposed Multi-Head LSTM Architecture}
\label{tab:architecture}
\begin{tabular}{@{}lllc@{}}
\toprule
\textbf{Component} & \textbf{Layer Type} & \textbf{Output Shape} \\ \midrule
\multirow{4}{*}{\textbf{Backbone}} & Input Sequence & (Batch, 20, 25)\\
 & LSTM (Layer 1) & (Batch, 20, 128) \\
 & LSTM (Layer 2) & (Batch, 20, 128) \\
 & LayerNorm + Dropout & (Batch, 128) \\ \midrule
\multirow{2}{*}{\textbf{Head 1 (Stress)}} & Dense + ReLU + Drop & (Batch, 32) \\
 & Dense (Logit) & (Batch, 1) \\ \midrule
\multirow{2}{*}{\textbf{Head 2 (Ineff)}} & Dense + ReLU + Drop & (Batch, 32) \\
 & Dense (Logit) & (Batch, 1) \\ \midrule
\multirow{2}{*}{\textbf{Head 3 (Power)}} & Dense + ReLU + Drop & (Batch, 32) \\
 & Dense (Linear) & (Batch, 3)\\ \bottomrule
\end{tabular}
\end{table}

\textbf{III) Multi-Task Optimization and Loss Functions:}
We define a composite loss function comprising three distinct, task-specific formulations. For the thermal stress prediction (Head 1), the dataset exhibits a severe class imbalance (16\% stressed vs 84\% normal). Standard cross entropy would heavily bias the model toward the majority class. Therefore, we optimize this head using Focal Loss \cite{lin2017focal}, which dynamically scales the loss based on prediction confidence, heavily penalizing misclassifications on hard, minority class examples while down weighting easy, majority class examples:
\begin{equation}
\mathcal{L}_{\text{stress}} = - \frac{1}{N} \sum_{j=1}^{N} \alpha_t (1 - \hat{p}_{\text{stress},j})^\gamma \log(\hat{p}_{\text{stress},j}),
\end{equation}
where $\hat{p}_{\text{stress},j}$ is the model's estimated probability for the ground truth class, the focusing parameter $\gamma$ dictates the rate at which easy examples are down weighted, and $\alpha_t$ provides explicit weighting for the positive class.

For the cooling inefficiency detection (Head 2), the class imbalance is moderate (33\% inefficient vs. 67\% efficient). We optimize this head using a Binary Cross-Entropy (BCE) loss with positive weight scaling to penalize false negatives:
\begin{equation}
\begin{split}
\mathcal{L}_{\text{ineff}} = - \frac{1}{N} \sum_{j=1}^{N} \Big[ w_{pos} y_{\text{ineff},j} \log(\hat{p}_{\text{ineff},j}) \\
+ (1 - y_{\text{ineff},j}) \log(1 - \hat{p}_{\text{ineff},j}) \Big],
\end{split}
\end{equation}
where $y_{\text{ineff},j} \in \{0,1\}$ is the true inefficiency label and $w_{pos}$ is calculated dynamically as the ratio of negative to positive samples in the training set.

For the power forecasting task (Head 3), we utilize the Mean Squared Error (MSE) loss over the normalized targets:
\begin{equation}
\mathcal{L}_{\text{power}} = \frac{1}{N} \sum_{j=1}^{N} \sum_{h=1}^{3} (\hat{y}_{\text{power},j}^{(h)} - y_{\text{power},j}^{(h)})^2,
\end{equation}
where $h$ indexes the three forecast horizons. 

Because the MSE targets are scaled to the $[0,1]$ range, the raw magnitude of $\mathcal{L}_{\text{power}}$ is exceptionally small compared to the BCE loss. Left unscaled, the gradients from Head 3 would be entirely eclipsed by Head 2. We formulate the final multitask objective as a weighted linear combination of the three losses:
\begin{equation}
\mathcal{L}_{\text{total}} = \lambda_1 \mathcal{L}_{\text{stress}} + \lambda_2 \mathcal{L}_{\text{ineff}} + \lambda_3 \mathcal{L}_{\text{power}}.
\end{equation}
Through empirical gradient observation during training, the scalar weights ($\lambda_1$, $\lambda_2$, $\lambda_3$) are tuned to ensure that the backpropagated errors from all three heads arrive at the shared LSTM backbone with comparable magnitudes, enabling stable convergence. The precise hyperparameter values utilized for these loss functions are detailed below.

\section{Experimental Setup}
\label{sec:experimental_setup}




We used the Air Cooled Tropical Data Center (TDC2.0) dataset \cite{N9/BLBQ2T_2023} as the basis for our experimental evaluation. After preprocessing and feature engineering, the final structured dataset includes 520,965 contiguous timesteps recorded every 30 seconds. This is equivalent to six months of continuous facility operation.


As determined in the feature selection phase, the input space comprises $N_f = 25$ variables. These features are organized into groups based on their purpose: environmental metrics (temperature and humidity of outdoor and supply air), electrical metrics (power draw of each component), derived weather-load deltas, and total system power. data over time. Table \ref{tab:dataset_feature_summary} shows a summary of the global features and the specific feature composition of the final dataset.

\begin{table}[htbp]
\centering
\caption{Processed Dataset Metrics and Feature Space Composition}
\label{tab:dataset_feature_summary}
\begin{tabular}{@{}llc@{}}
\toprule
\textbf{Category} & \textbf{Description} & \textbf{Count} \\ \midrule
\multicolumn{3}{c}{\textbf{Global Dataset Characteristics}} \\ \midrule
Total Duration & Six months (Sept 2022 to Feb 2023) & -- \\
Observations & 520,965 contiguous timesteps & -- \\
Resolution & 30 seconds per timestep & -- \\ \midrule
\multicolumn{3}{c}{\textbf{Input Feature Space ($N_f = 25$)}} \\ \midrule
Environmental & OATH and SATH (Temp and Humidity) & 8 \\
Electrical & PCM, PCP1, PCP2, PEF, PCF power & 10 \\
Derived Deltas & Weather load (Temp and Humidity) & 4 \\
System Totals & Individual PCU and facility total power & 3 \\ \bottomrule
\end{tabular}
\end{table}

\subsubsection{Phase-Aware Chronological Splitting}


Standard random splitting strategies are unsuitable for time-series forecasting because they introduce temporal leakage. While conventional temporal splitting preserves chronological order, applying it directly to the six-month dataset would separate distinct operational phases of the testbed into different partitions. As these phases exhibit different operating characteristics,such a split could introduce distributional shifts between the training and testing sets, limiting the model's ability to learn representative cooling-system behaviors. To fix this issue, a phase-aware chronological split is used. The dataset was split into its four main operational phases: controlling the temperature, controlling the humidity, controlling the airflow, and long-term reliability. In strict chronological order, each phase was split into three sets: 70\% for training, 15\% for testing, and 15\% for validation.
\subsubsection{Training Details and Hyperparameter Configuration}


The proposed multihead LSTM model was developed based on the PyTorch deep learning platform. In order to optimize the network parameters, we used the Adam optimizer with the initial learning rate of 3 x 10-4 and weight decay of 1 x 10-4 to add L2 regularization. 

Learning rate scheduling was controlled through a One Cycle policy in order to speed up convergence and ensure that the network does not fall into local minima. This method slowly raises the learning rate for the first 20\% of training and then lowers it again in a way that follows a cosine curve. A total of 200 epochs were used to train the model with a batch size of 1024. To stop exploding gradients, which is especially important when training recurrent architectures on long sequences, gradient clipping was used with a maximum norm of 1.0. Table \ref{tab:hyperparameters} shows a summary of all the architectural, loss, and training hyperparameters.

\begin{table}[htbp]
\centering
\caption{Optimization and Training Hyperparameters}
\label{tab:hyperparameters}
\begin{tabular}{@{}llc@{}}
\toprule
\textbf{Component} & \textbf{Hyperparameter} & \textbf{Value} \\ \midrule
\multicolumn{3}{c}{\textbf{Loss Function Weights}} \\ \midrule
Head 1 (Stress) & Focal Loss Alpha ($\alpha_t$) & 0.75 \\
Head 1 (Stress) & Focal Loss Gamma ($\gamma$) & 3.0 \\
Objective Function & Head 1 Scalar Weight ($\lambda_1$) & 5.0 \\
Objective Function & Head 2 Scalar Weight ($\lambda_2$) & 1.0 \\
Objective Function & Head 3 Scalar Weight ($\lambda_3$) & 200.0 \\ \midrule
\multicolumn{3}{c}{\textbf{Global Training Parameters}} \\ \midrule
Optimization & Initial Learning Rate & $3 \times 10^{-4}$ \\
Optimization & Weight Decay & $1 \times 10^{-4}$ \\
Optimization & Batch Size & 1024 \\
Optimization & Total Epochs & 200 \\ \bottomrule
\end{tabular}
\end{table}

We implemented a functional prototype of the proposed
server-room monitoring system using RTI Connext DDS
7.3.0 and the RTI Connext Python API. The prototype
consists of three distributed sensing nodes representing
temperature, humidity, and smart-meter sensors, and a
central server that subscribes to all topics and displays the
received measurements in real time.
To evaluate communication performance, latency and
throughput experiments were conducted over a wireless network
using two machines connected through a TP-Link router
with a nominal bandwidth of 150 Mbps. One machine acted
as the sensor-side publisher, while the other acted as the
server-side subscriber. The publisher ran on a Windows 10 machine with an Intel Pentium dual-core 2.4 GHz CPU, 12 GB RAM, and a Ralink RT3290 802.11n adapter; the subscriber ran on Windows 10 with an Intel Core i5, 16 GB RAM, and an Intel Dual Band Wireless-AC 7265 802.11n adapter. Both adapters operated at 2.4 GHz with a negotiated link speed of 150 Mbps. In the latency experiment, the publisher transmitted 100
samples, each with a payload size of 128 bytes, under reliable and best-effort QoS settings. The reliable configuration
represents the target configuration of the proposed monitoring
system, while the best-effort configuration was used as a
comparison baseline. To avoid clock-synchronization errors,
latency was measured using an echo-based round-trip method.
In this method, the sensor-side publisher sends a message
to the server-side subscriber, and the subscriber immediately echoes the same message back. The publisher then measures
the round-trip time (RTT) using its local clock. Fig.~\ref{fig:latency_samples}  shows
the estimated one-way latency for each sample, computed as
half of the measured RTT. This procedure was repeated 100
times for each QoS configuration. 

\section{Performance Evaluation}\label{sec:results}

 The evaluation is systematically organized into two primary domains to assess both the communication infrastructure and the predictive analytics engine. First, we analyze the communication performance of the brokerless DDS middleware, specifically evaluating the network latency, throughput, and the effectiveness of the applied QoS policies in ensuring reliable sensor data transmission. Second, we rigorously evaluate the predictive capabilities of the multi-head LSTM architecture across its three distinct tasks.

\subsection{DDS Performance Evaluation}


The experimental results indicate that DDS achieved low latency and high throughput under both QoS modes in the tested wireless environment. However, the two QoS profiles exhibit different trade-offs in terms of delay, tail behavior, and throughput efficiency.

\begin{figure}[htbp]
\centering
\includegraphics[width=\linewidth]{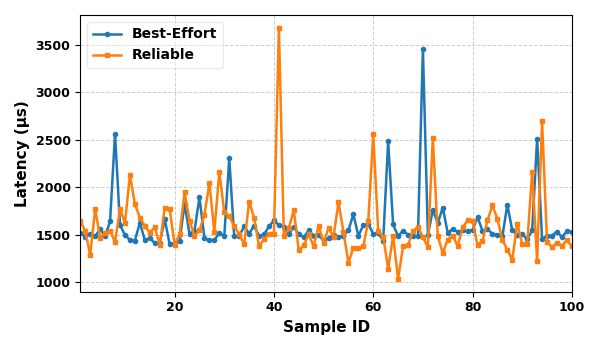}
\caption{Sample-by-sample one-way latency comparison between Best-Effort and Reliable QoS profiles for 100 transmitted samples of 128-byte messages.}
\label{fig:latency_samples}
\end{figure}

The sample-by-sample latency behavior is shown in Fig.~\ref{fig:latency_samples}, where both QoS profiles remain clustered around the low-millisecond range for most of the 100 transmitted samples. Best-Effort generally exhibits slightly lower and more stable latency, whereas Reliable shows a similar baseline level but with more spikes at a few measured samples. This indicates that the latency cost of reliability is expressed through occasional higher-delay events.

\begin{figure}[!t]
\centering
\includegraphics[width=\linewidth]{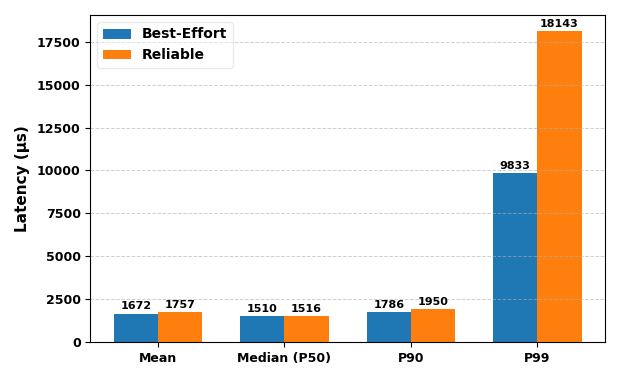}
\caption{Summary latency comparison between Best-Effort and Reliable QoS profiles in terms of mean, median (P50), P90, and P99 one-way latency for 128-byte messages.}
\label{fig:latency_summary}
\end{figure}

This trend becomes clearer in the summary statistics presented in Fig.~\ref{fig:latency_summary}. Under Reliable QoS, the average one-way latency was 1.757 ms and the median latency was 1.516 ms. Under Best-Effort QoS, the average latency was slightly lower at 1.672 ms, while the median latency was 1.510 ms. The near-equality of the medians indicates that the two QoS profiles have almost identical typical latency under normal operating conditions. 
 Fig.~\ref{fig:latency_summary} shows that Reliable QoS introduces a heavier latency tail. The P90 latency increases from 1.786 ms under Best-Effort to 1.950 ms under Reliable, while the P99 latency rises much more sharply, from 9.833 ms to 18.143 ms. This means that the main overhead of Reliable QoS is not a large shift in average or median latency, but a greater likelihood of occasional delay spikes.

\begin{figure}[!t]
\centering
\includegraphics[width=\linewidth]{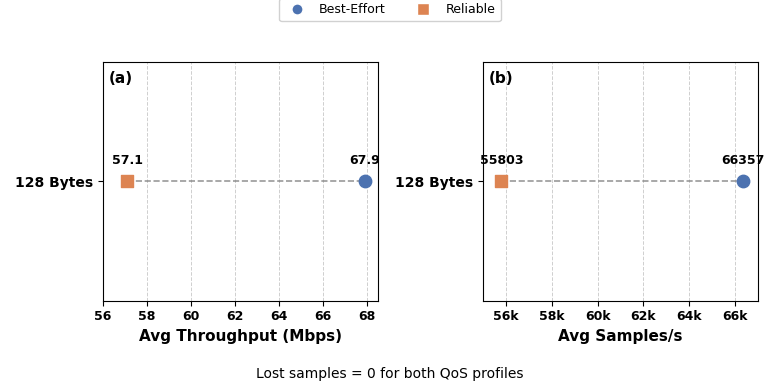}
\caption{Throughput comparison between Best-Effort and Reliable QoS profiles for 128-byte messages in terms of average throughput (Mbps) and average transmission rate (samples/s). No packet loss was observed in either mode.}
\label{fig:throughput_summary}
\end{figure}

The throughput results, summarized in Fig.~\ref{fig:throughput_summary}, show the opposite side of this trade-off. Best-Effort achieved 67.9 Mbps and 66,357 samples/s, whereas Reliable achieved 57.1 Mbps and 55,803 samples/s. No lost samples were recorded for either profile during the 100-sample experiment. These results confirm that Best-Effort is more efficient in terms of pure transmission rate, which is expected because it avoids the extra control traffic associated with guaranteed delivery. Reliable, by contrast, consumes additional communication and middleware resources to maintain delivery assurance, which reduces the effective throughput.


Although Best-Effort produced slightly lower average latency and higher throughput, Reliable QoS remains the more appropriate choice for the proposed server-room monitoring system. The target workload is periodic monitoring traffic, not a high-rate stream in which minor loss is easily tolerated. Measurements such as temperature, humidity, and power consumption are operationally important because missed abnormal readings may delay fault detection or weaken safety assessment. In this context, the system benefits more from dependable delivery than from minimizing latency by a small margin. The results support this design decision: the median latency difference between the two modes is negligible, while the Reliable profile still maintains a latency level in the low-millisecond range, which is acceptable for periodic infrastructure monitoring. A recent study \cite{gambo2025efficient}   demonstrates the DDS reliability in transmitting time-sensitive data.


\subsection{Multi-Head LSTM Performance Evaluation}


To evaluate the generalization capability of the proposed architecture on unseen data, the model was tested with the 15\% chronological testing set that was set aside during the phase-aware splitting stage. The evaluation is organized around the three different projection heads to see how well the shared recurrent backbone captured the system dynamics that were related to each other. To evaluate performance on the binary classification tasks of finding thermal stress and cooling inefficiency, we use precision, recall, and the F1-score to get a balanced picture that takes into account the class imbalances. Additionally, confusion matrices and Receiver Operating Characteristic (ROC) curves are employed to analyze the classification behavior comprehensively. For the continuous power forecasting task, we use MSE to measure the predictive error across the three chosen future horizons at the 30-second sensor frequency.

\begin{figure}[!t]
    \centering
    \includegraphics[
        width=\columnwidth,
        trim=0 0 0 10,
        clip
    ]{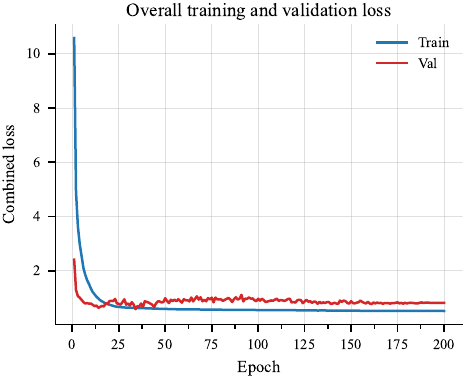}
    \caption{Overall training and validation loss across 200 epochs.}
    \label{fig:overall_loss}
\end{figure}

\begin{figure}[!t]
    \centering
    \includegraphics[
        width=\columnwidth,
        trim=0 0 0 10,
        clip
    ]{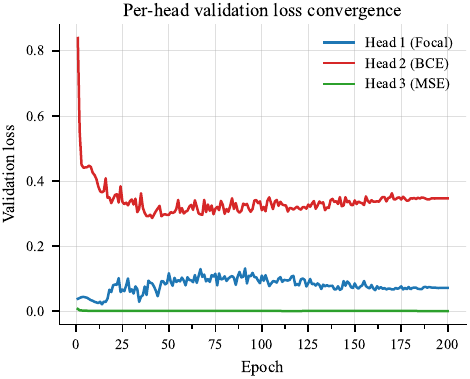}
    \caption{Comparison of the validation loss across the three individual heads.}
    \label{fig:heads_val_loss}
\end{figure}

\subsubsection{Multi-Task Training Loss}

We looked at the training dynamics of the proposed multi-head LSTM architecture over 200 epochs to see if it converged and to see if the shared recurrent backbone could learn the interrelated physical dynamics without overfitting too much. Figure \ref{fig:overall_loss} shows how the total training and validation loss has changed over time. The validation loss closely follows the training path during the optimization process and steadily goes down until it levels off in the later epochs. This close alignment indicates that the regularization methods used, such as dropout and weight decay, helped reduce overfitting and let the model keep generalizing on the new chronological validation split.

One of the biggest problems with multi-task learning is making sure that the gradient of one task doesn't take over the others during backpropagation. To solve this problem, the total objective function used scaled scalar weights ($\lambda_1, \lambda_2, \lambda_3$) to balance the different loss formulations of the classification and regression heads. Figure \ref{fig:heads_val_loss} shows the validation loss for each projection head in detail. The simultaneous and relatively stable convergence of the Focal Loss (Head 1), Binary Cross-Entropy Loss (Head 2), and Mean Squared Error (Head 3) indicates that this weighting strategy was effective. This means that the shared LSTM layers were able to learn general representations that helped them predict thermal stress, cooling inefficiency, and future power usage at the same time.

\subsubsection{Thermal Stress and Inefficiency Classification}

The initial phase of the evaluation concentrates on the two distinct classification tasks. When it comes to critical facility infrastructure, relying only on raw accuracy can be misleading because the operational data has an inherent class imbalance. For example, extreme thermal stress is an uncommon occurrence in contrast to normal operation. So, we use Accuracy, Precision, Recall, and the F1-score to see how well the model works. Precision tells us how many of the predicted positive events were actually correct, and Recall tells us how well the model can find all of the true positive events in the dataset. The F1-score is the harmonic mean of Precision and Recall. It gives you one number that balances false positives and false negatives.

Table \ref{tab:classification_metrics} summarizes the final performance metrics achieved on the holdout testing set for both classification heads.

\begin{table}[htbp]
\centering
\caption{Classification Performance Metrics}
\label{tab:classification_metrics}
\begin{tabular}{@{}lcccc@{}}
\toprule
\textbf{Task} & \textbf{Accuracy} & \textbf{Precision} & \textbf{Recall} & \textbf{F1-Score} \\ \midrule
Head 1 (Thermal Stress) & 0.9618 & 0.9269 & 0.9167 & 0.9217 \\
Head 2 (Inefficiency) & 0.6920 & 0.6115 & 0.9131 & 0.7325 \\ \bottomrule
\end{tabular}
\end{table}


\begin{figure}[htbp]
    \centering
    \includegraphics[width=\columnwidth]{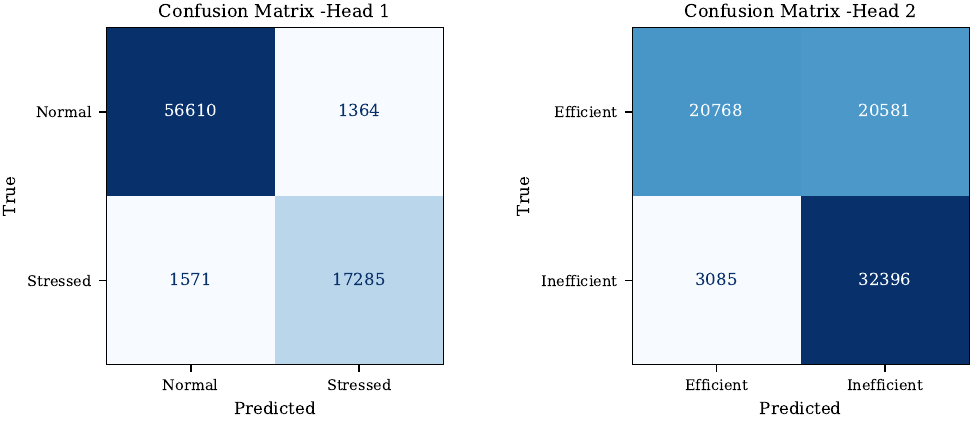}
    \caption{Confusion matrices evaluated on the testing dataset. The left matrix displays the performance of the thermal stress classification, and the right matrix illustrates the cooling inefficiency detection.}
    \label{fig:confusion_matrices}
\end{figure}

The results show that the two tasks had different levels of physical complexity, which led to different levels of performance. The model was 96.18\% accurate and had an F1 score of 0.9217 for predicting thermal stress (Head 1). The recall value of 0.9167 is pretty good, which is very important for this safety task because it means the system can catch most unsafe temperature events before they happen. 

The cooling inefficiency detection (Head 2), on the other hand, is a more difficult predictive task. The model had a high recall of 0.9131, indicating that it is effective in identifying most of the periods of wasteful use of energy. Nonetheless, the precision is considerably lower at 0.6115, which reduces the total accuracy to 0.6920. This implies that although the model is very sensitive to inefficiency, it over predicts inefficiency hence resulting in a number of false positive warnings. This is common in anomaly detection where the decision threshold is optimized (here to 0.378) to focus on capturing true positive events rather than ensuring perfect precision.

 Figure \ref{fig:confusion_matrices} presents the confusion matrices for both classification tasks, providing further insight into the model's predictive behavior. In critical infrastructure monitoring, false negatives are particularly critical because they represent missed warnings that may result in equipment damage, whereas false positives primarily introduce additional administrative overhead. The left panel (Head 1, thermal stress) shows predictions concentrated along the true positive and true negative diagonal, with few false negatives, indicating reliable detection of stressed operating conditions. The right panel (Head 2, cooling inefficiency) helps explain the previously observed lower precision score. Although the model correctly identifies most inefficient operating windows, the relatively higher number of false positives reflects the conservative decision threshold, which prioritizes the detection of potential energy waste at the expense of a higher false-alarm rate.



In order to finish the assessment of the classification tasks, ROC curves are examined.

\begin{figure}[htbp]
    \centering
    \includegraphics[width=\columnwidth]{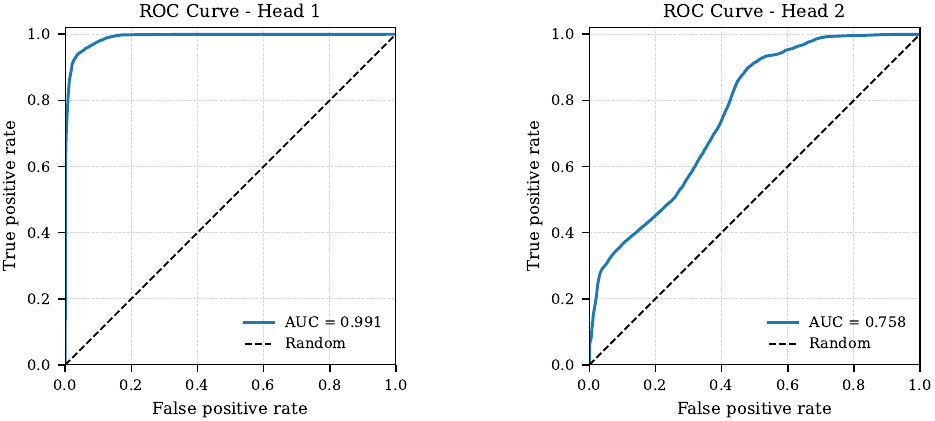}
    \caption{ROC curves for the classification tasks evaluated on the test set. The left panel shows the thermal stress prediction, and the right panel illustrates the cooling inefficiency detection.}
    \label{fig:roc_curves}
\end{figure}
The ROC curves of the two prediction heads are shown in Fig \ref{fig:roc_curves}. The left panel of the figure shows the high level of discrimination of the model in predicting thermal stress with an AUC of 0.99. The curve is steeply increasing to the upper left, which means that the network can achieve a high true positive rate, and keep the false alarms to a minimum over a large set of possible thresholds.

The right panel of Fig \ref{fig:roc_curves}, which depicts the cooling inefficiency detection, has an AUC of 0.76. This reduced separability in comparison with the thermal stress head is not surprising. The detection of systemic cooling inefficiencies using latent temporal correlations of power and environmental measurements is inherently less certain than the detection of direct temperature violations. 

\subsubsection{Continuous Power Forecasting}
 Mean Absolute Error (MAE), Root Mean Squared Error (RMSE), Coefficient of Determination ($R^2$), and Mean Absolute Percentage Error (MAPE) are used to assess the performance based on the three predictive horizons defined ($t+5$, $t+10$, and $t+30$ minutes).

Table \ref{tab:regression_metrics} summarizes the overall forecasting performance across the entire testing dataset.

\begin{table}[htbp]
\centering
\caption{Power Forecasting Error Metrics on the Complete Testing Set}
\label{tab:regression_metrics}
\begin{tabular}{@{}lcccc@{}}
\toprule
\textbf{Horizon} & \textbf{MAE (W)} & \textbf{RMSE (W)} & \textbf{$R^2$} & \textbf{MAPE (\%)} \\ \midrule
$t+5$ min  & 1444.6 & 2236.1 & 0.8558 & 7.82 \\
$t+10$ min & 1426.7 & 2256.9 & 0.8531 & 7.70 \\
$t+30$ min & 1485.8 & 2380.1 & 0.8365 & 8.00 \\ \bottomrule
\end{tabular}
\end{table}

\begin{figure*}[htbp]
    \centering

    \begin{minipage}{0.32\textwidth}
        \centering
        \includegraphics[width=\linewidth]{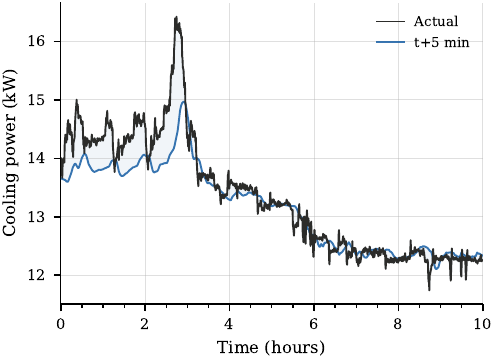}

        \small{(a) 5-Minute Forecast Horizon ($t+5$)}
    \end{minipage}
    \hfill
    \begin{minipage}{0.32\textwidth}
        \centering
        \includegraphics[width=\linewidth]{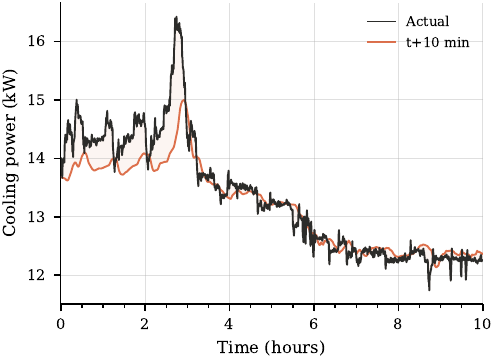}

        \small{(b) 10-Minute Forecast Horizon ($t+10$)}
    \end{minipage}
    \hfill
    \begin{minipage}{0.32\textwidth}
        \centering
        \includegraphics[width=\linewidth]{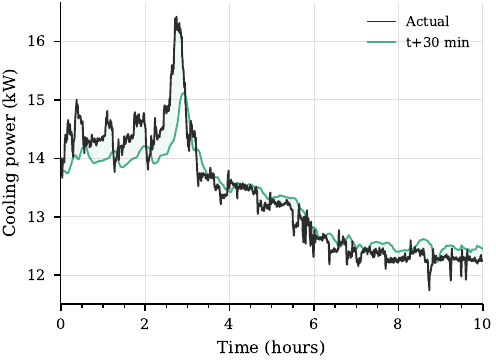}

        \small{(c) 30-Minute Forecast Horizon ($t+30$)}
    \end{minipage}

    \caption{Time-series comparison of actual versus predicted cooling power consumption across three future horizons during a dynamic 10-hour window.}
    \label{fig:forecast_horizons}
\end{figure*}

The findings show that the common recurrent backbone retrieved very significant temporal features to predict power. The model had an $R^2$ score of over 0.83 in all horizons, indicating that it is able to explain a substantial amount of the variation in the cooling power consumption. It is noteworthy that the error measures do not change significantly with the various horizons, which indicates that the model is not sensitive to quick degradation when making predictions that are further into the future.
\begin{figure*}[htbp]
    \centering
    \includegraphics[width=0.8\textwidth]{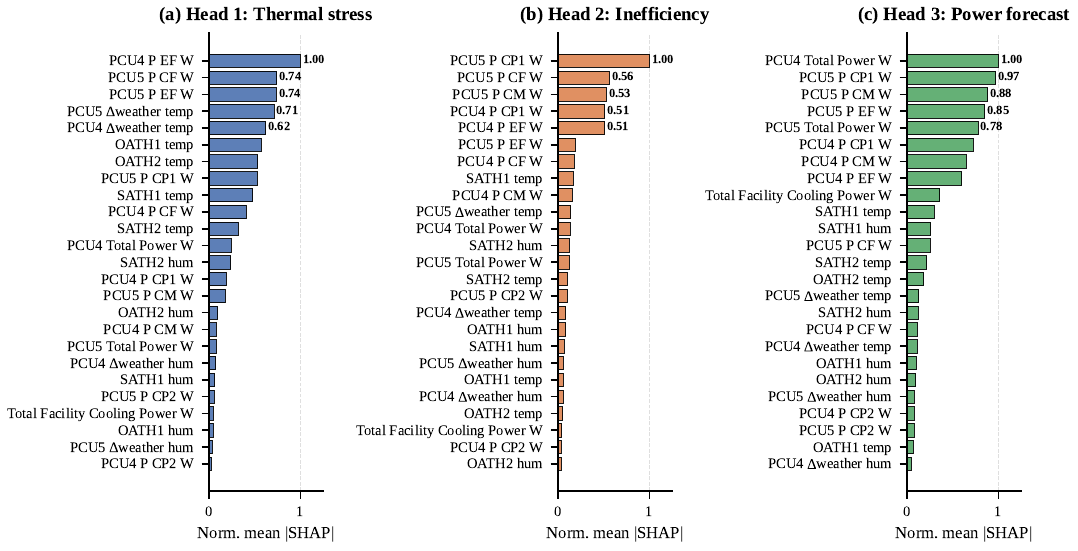}
    \caption{Global SHAP feature importance for the three predictive tasks.}
    \label{fig:shap_importance}
\end{figure*} 

In order to visualize the predictive behavior, Fig. \ref{fig:forecast_horizons} compares the predicted power with the actual ground truth over a typical 10-hour period with dynamic workload changes. The time-series plots show that the prediction horizon and the capability of the model to follow sudden electrical transients are correlated. Fig. \ref{fig:forecast_horizons}(a) shows that the forecast at $t+5$ is very close to the ground truth, reflecting both macro-level trends and spikes in power consumption at higher frequencies, which leads to low MAPE of 5.27\% in this particular window. The forecast automatically becomes less accurate (MAPE rises to 7.32\%), as the horizon is extended to $t+30$ as shown in panel (c) of the figure. Although the model still effectively predicts larger scale operational cycles, the prediction curve is much smoother, and tends to underestimate the absolute peak amplitude of sudden spikes. This is expected to happen in physical system forecasting, where instantaneous electrical bursts caused by stochastic bursts of work in the short term are naturally hard to predict 30 minutes ahead.

\subsubsection{Model Interpretability and Feature Importance}
 Figure \ref{fig:shap_importance} illustrates the normalized global feature importance for the three multi-task objectives. 
 The way the importance scores are spread out is very similar to how thermodynamics works in cooling a data center. The model gives the most weight to the power use of the evaporator and condenser fans (PCU4-P-EF, PCU5-P-CF) and the temperature outside when predicting thermal stress (Fig. \ref{fig:shap_importance}a). This suggests the model has learned that thermal stress is strongly correlated with periods where the cooling units are operating their fans at peak capacity to combat high external heat loads.

On the other hand, compressor power metrics (like PCU5-P-CP1) and total unit power make up most of the cooling inefficiency detection (Fig. \ref{fig:shap_importance}b). This makes sense from a physical point of view because mechanical cooling inefficiency is mostly caused by compressors using too much electricity compared to the amount of heat they are removing from the environment. Lastly, the power forecasting head (Fig. \ref{fig:shap_importance}c) uses a mix of the current power states of the compressor and fan to guess how much energy will be needed in the future. The SHAP analysis shows that the shared LSTM backbone was able to learn how to find and use the right physical parameters for each monitoring task.

\section{CONCLUSION}

This paper presented a granular, distributed monitoring architecture aimed at supporting the operational safety and energy efficiency of modern server rooms. By decoupling the edge sensing layer from the centralized analytics engine using a brokerless DDS middleware, the system was designed to reliably transmit component-level environmental and electrical metrics. Our evaluation of this communication layer indicated a stable throughput of approximately 57.1 Mbps alongside an average end-to-end latency of 1.757 milliseconds under Reliable QoS, suggesting that the middleware can comfortably support the required periodic sensor polling intervals without creating bottlenecks. To analyze the complex temporal dynamics of this environment, a multi-head LSTM network was developed and evaluated on the empirical TDC2.0 dataset. The experimental observations indicate that a single shared recurrent backbone, balanced via task-specific loss weighting, can learn representations helpful for simultaneous multi-task prediction. The model showed promising results in classifying critical thermal stress events, achieving an F1-score of 0.92, and identifying systemic cooling inefficiencies.

Although the proposed framework demonstrated promising performance, its evaluation was conducted using the publicly available TDC2.0 dataset. While this dataset spans six months and captures diverse operating conditions, further validation using additional datasets and deployment environments would strengthen the assessment of the framework's generalizability. Future work will therefore investigate the proposed approach across different server-room configurations.

\section*{Acknowledgment}
The authors would like to acknowledge all the support provided by King Fahd University of Petroleum \& Minerals (KFUPM).


%





\ifCLASSOPTIONcaptionsoff
  \newpage
\fi





\bibliographystyle{IEEEtran}
\bibliography{IEEEabrv,Bibliography}

@data{N9/BLBQ2T_2023,
author = {Le, Van Duc},
publisher = {DR-NTU (Data)},
title = {{An Air-Cooled Tropical Data Center (TDC2.0) Dataset}},
year = {2023},
version = {V2},
doi = {10.21979/N9/BLBQ2T},
url = {https://doi.org/10.21979/N9/BLBQ2T}
}

@inproceedings{lin2017focal,
  title={Focal loss for dense object detection},
  author={Lin, Tsung-Yi and Goyal, Priya and Girshick, Ross and He, Kaiming and Doll{\'a}r, Piotr},
  booktitle={Proceedings of the IEEE international conference on computer vision},
  pages={2980--2988},
  year={2017}
}

@article{wang2013sensor,
  author    = {X. Wang and X. Wang and G. Xing and J. Chen and C.-X. Lin and Y. Chen},
  title     = {Intelligent Sensor Placement for Hot Server Detection in Data Centers},
  journal   = {IEEE Transactions on Parallel and Distributed Systems},
  volume    = {24},
  number    = {8},
  pages     = {1577--1588},
  year      = {2013},
  doi       = {10.1109/TPDS.2012.254}
}

@inproceedings{mousavi2015energy,
  author    = {A. Mousavi and V. Vyatkin and Y. Berezovskaya and X. Zhang},
  title     = {Towards Energy Smart Data Centers: Simulation of Server Room Cooling System},
  booktitle = {Proceedings of the 2015 IEEE 20th International Conference on Emerging Technologies and Factory Automation (ETFA)},
  address   = {Luxembourg},
  pages     = {1--6},
  year      = {2015},
  doi       = {10.1109/ETFA.2015.7301573}
}

@inproceedings{purwanto2018fuzzy,
  author    = {F. H. Purwanto and E. Utami and E. Pramono},
  title     = {Design of Server Room Temperature and Humidity Control System Using Fuzzy Logic Based on Microcontroller},
  booktitle = {Proceedings of the 2018 International Conference on Information and Communications Technology (ICOIACT)},
  pages     = {390--395},
  year      = {2018},
  doi       = {10.1109/ICOIACT.2018.8350770}
}

@inproceedings{onibonoje2019iot,
  author    = {M. O. Onibonoje and P. N. Bokoro and N. I. Nwulu and S. L. Gbadamosi},
  title     = {An IoT-Based Approach to Real-Time Conditioning and Control in a Server Room},
  booktitle = {2019 International Conference on Artificial Intelligence and Data Processing Symposium (IDAP)},
  pages     = {1--6},
  year      = {2019},
  doi       = {10.1109/IDAP.2019.8875880}
}

@article{li2020drl,
  author    = {Y. Li and Y. Wen and D. Tao and K. Guan},
  title     = {Transforming Cooling Optimization for Green Data Center via Deep Reinforcement Learning},
  journal   = {IEEE Transactions on Cybernetics},
  volume    = {50},
  number    = {5},
  pages     = {2002--2013},
  year      = {2020},
  doi       = {10.1109/TCYB.2019.2927410}
}

@article{zhang2021survey,
  author    = {Q. Zhang and others},
  title     = {A Survey on Data Center Cooling Systems: Technology, Power Consumption Modeling and Control Strategy Optimization},
  journal   = {Journal of Systems Architecture},
  volume    = {119},
  pages     = {102253},
  year      = {2021},
  doi       = {10.1016/j.sysarc.2021.102253}
}

@article{chen2023mlcooling,
  author    = {X. Chen and R. Tu and X. Yang},
  title     = {Parameter Prediction Optimization of Data Center’s Heat Dissipation System Using Machine Learning Algorithms},
  journal   = {Applied Thermal Engineering},
  volume    = {232},
  pages     = {121047},
  year      = {2023},
  doi       = {10.1016/j.applthermaleng.2023.121047}
}

@article{li2024physicsann,
  author    = {X. Li and X. Wang and Z. He and X. Chen and Z. Li},
  title     = {Combining Physical Laws and ANN for Predicting Energy Consumption of Data Center Cooling Systems},
  journal   = {Energy and Buildings},
  volume    = {311},
  pages     = {114170},
  year      = {2024},
  doi       = {10.1016/j.enbuild.2024.114170}
}

@article{kula2025mltemp,
  author    = {A. Kula and D. D{\k{a}}browski and M. Blachnik and M. Sajkowski and A. Smalcerz and Z. Kami{\'n}ski},
  title     = {Modelling the Temperature of a Data Centre Cooling System Using Machine Learning Methods},
  journal   = {Energies},
  volume    = {18},
  number    = {10},
  pages     = {2581},
  year      = {2025},
  doi       = {10.3390/en18102581}
}

@article{shuaibu2026smart,
  author    = {A. S. Shuaibu and S. U. Haq and B. Almadani and F. Aliyu and E. G. Al-Nahari},
  title     = {Smart Metering Meets AI: Real-Time Appliance Monitoring and Anomaly Detection over DDS Middleware},
  journal   = {IEEE Transactions on Industry Applications},
  volume    = {62},
  number    = {2},
  pages     = {2326--2342},
  year      = {2026},
  doi       = {10.1109/TIA.2025.3608687}
}

@online{iea_ai_energy_2024,
  author = {{IEA}},
  title = {What the data centre and AI boom could mean for the energy sector},
  year = {2024},
  url = {https://www.iea.org/commentaries/what-the-data-centre-and-ai-boom-could-mean-for-the-energy-sector},
  note = {Accessed: 2026-04-25}
}

@online{iea_energy_ai_report,
  author = {{International Energy Agency (IEA)}},
  title = {Energy and AI: Energy demand from AI},
  year = {2024},
  url = {https://www.iea.org/reports/energy-and-ai/energy-demand-from-ai},
  note = {Accessed: 2026-04-25}
}

@online{doe_cooling_2023,
  author = {{U.S. Department of Energy}},
  title = {DOE Announces \$40 Million for More Efficient Cooling in Data Centers},
  year = {2023},
  url = {https://www.energy.gov/articles/doe-announces-40-million-more-efficient-cooling-data-centers},
  note = {Accessed: 2026-04-25}
}

@online{reuters_datacenter_cooling_2025,
  author = {{Reuters}},
  title = {Keeping cool: Heat is key challenge for data centers in AI era},
  year = {2025},
  url = {https://www.reuters.com/business/keeping-cool-heat-key-challenge-data-centers-ai-2025-11-28/},
  note = {Accessed: 2026-04-25}
}

@article{gambo2025efficient,
  title={An Efficient Framework for Automated Cyber Threat Intelligence Sharing},
  author={Gambo, Muhammad Dikko and Khan, Ayaz H and Almulhem, Ahmad and Almadani, Basem},
  journal={Electronics},
  volume={14},
  number={20},
  pages={4045},
  year={2025},
  publisher={MDPI}
}

@techreport{lbl_server_data_2021,
  author = {{Lawrence Berkeley National Laboratory}},
  title = {Accessing Onboard Server Data},
  year = {2021},
  url = {https://datacenters.lbl.gov/sites/default/files/FINAL%20Accessing%20Onboard%20Server%20Data%209-20-2021%20%281%29.pdf},
  note = {Accessed: 2026-04-25}
}

@techreport{ashrae_thermal_guidelines,
  author = {{ASHRAE}},
  title = {Thermal Guidelines for Data Centers: Past, Present and Future},
  year = {2025},
  url = {https://dallas-ashrae.org/images/meeting/041625/the_ashrae_thermal_guidelines_for_data_centers_____past__present_and_future.pdf},
  note = {Accessed: 2026-04-25}
}

@techreport{google_datacenter_efficiency,
  author = {{Google Research}},
  title = {Energy Efficiency in Data Centers (Google Whitepaper)},
  year = {2012},
  url = {https://static.googleusercontent.com/media/research.google.com/en//pubs/archive/42542.pdf},
  note = {Accessed: 2026-04-25}
}

@inproceedings{anwar2017monitor,
  author = {Anwar, H. and Santoso, H. and Khameswara, T. D. and Priantoro, A. U.},
  title = {Monitor-PoP ISP's PoP Room Temperature and Humidity Web-Based Monitoring Using Microcontroller},
  booktitle = {2017 IEEE 8th Control and System Graduate Research Colloquium (ICSGRC)},
  pages = {212--216},
  year = {2017},
  organization = {IEEE}
}

@inproceedings{yumang2016zigbee,
  author = {Yumang, A. N. and Paglinawan, C. C. and Sejera, M. M. and Lazam, A. S. and Pagtakhan, J. C. and Santos, J. S. B.},
  title = {ZigBee-Based Monitoring of Temperature and Humidity of Server Rooms Using Thermal Imaging},
  booktitle = {2016 6th IEEE International Conference on Control System, Computing and Engineering (ICCSCE)},
  pages = {452--454},
  year = {2016},
  organization = {IEEE}
}

@inproceedings{rani2024mqtt,
  author = {Rani, M. and Thimmaraj, K. and Sasikala, S. J.},
  title = {Live Temperature Monitoring System Using MQTT},
  booktitle = {2024 International Conference on Intelligent and Innovative Technologies in Computing, Electrical and Electronics (IITCEE)},
  pages = {1--6},
  year = {2024},
  organization = {IEEE}
}

@techreport{IEA2025EnergyAI,
  author      = {{International Energy Agency}},
  title       = {Energy and AI},
  institution = {International Energy Agency},
  type        = {World Energy Outlook Special Report},
  address     = {Paris},
  year        = {2025},
  month       = apr,
  url         = {https://www.iea.org/reports/energy-and-ai},
  note        = {Licence: CC BY 4.0}
}

@ARTICLE{9111678,
  author={Fabrício, Marcos Aurélio and Behrens, Frank Herman and Bianchini, David},
  journal={IEEE Latin America Transactions}, 
  title={Monitoring of Industrial Electrical Equipment using IoT}, 
  year={2020},
  volume={18},
  number={08},
  pages={1425-1432},
  doi={10.1109/TLA.2020.9111678}}

\vfill


\end{document}